\documentclass[letterpaper]{article} 
\usepackage{aaai2026}  
\usepackage{times}  
\usepackage{helvet}  
\usepackage{courier}  
\usepackage[hyphens]{url}  
\usepackage{graphicx} 
\usepackage{natbib}  
\usepackage{caption} 
\usepackage{algorithm}
\usepackage{algorithmic}

\usepackage{amsmath}
\usepackage{amsfonts}
\usepackage{amssymb}

\DeclareMathOperator*{\expect}{\mathbb{E}}

\usepackage{newfloat}
\usepackage{listings}
\DeclareCaptionStyle{ruled}{labelfont=normalfont,labelsep=colon,strut=off} 
\floatstyle{ruled}
\newfloat{listing}{tb}{lst}{}
\floatname{listing}{Listing}
\title{Learning When to Cooperate Under Heterogeneous Goals}

\author {
    Max Taylor-Davies\textsuperscript{\rm 1},
    Neil Bramley\textsuperscript{\rm 2},
    Christopher G. Lucas\textsuperscript{\rm 1}
}
\affiliations {
    \textsuperscript{\rm 1}School of Informatics, The University of Edinburgh\\
    \textsuperscript{\rm 2}Department of Psychology, The University of Edinburgh\\
    maxtaylordavies@gmail.com, 
}

\begin{document}

\maketitle

\begin{abstract}
A significant element of human cooperative intelligence lies in our ability to identify opportunities for fruitful collaboration; and conversely to recognise when the task at hand is better pursued alone. Research on flexible cooperation in machines has left this meta-level problem largely unexplored, despite its importance for successful collaboration in heterogeneous open-ended environments. Here, we extend the typical Ad Hoc Teamwork (AHT) setting to incorporate the idea of agents having heterogeneous goals that in any given scenario may or may not overlap. We introduce a novel approach to learning policies in this setting, based on a hierarchical combination of imitation and reinforcement learning, and show that it outperforms baseline methods across extended versions of two cooperative environments. We also investigate the contribution of an auxiliary component that learns to model teammates by predicting their actions, finding that its effect on performance is inversely related to the amount of observable information about teammate goals. 
\end{abstract}


\section{Introduction}
The human capacity for cooperation and collaborative problem-solving is one of our most remarkable adaptations, having arguably played a crucial role in enabling us to develop culture and institutions that extend across space and time \cite{tomasello2012two, henrich2018secret}. It is thus no surprise that the creation of autonomous systems with human-like cooperative abilities has been a longstanding goal of research in AI and robotics. Most recently, work under the umbrella of `ad hoc teamwork' (AHT) has sought to develop agents that can collaborate `on-the-fly' with previously unseen teammates. Despite impressive progress, the AHT setting as typically considered leaves aside certain aspects that are crucial to real-world collaboration. In particular, AHT tends to assume that every scenario is equally cooperative; i.e. in any given `episode', it is \emph{always} optimal for the AHT agent to pursue some form of collaboration with the other agents in their environment. In the real world, things are not so simple---while some scenarios present fruitful opportunities for collaboration, in others it makes more sense to take an independent approach and focus on whatever we can achieve alone. For example, it makes sense to share travel with a friend if you have the same destination, but not if you're going in opposite directions. A truly human-like collaborator should be able to distinguish between these settings and adapt their behaviour accordingly. 

While the definition of AHT explicitly allows for agents to have different reward functions \cite{mirsky2022survey}, this has been little explored in prior work. We focus on the setting where agents have broadly the same \emph{underlying} goal (e.g. collect fruits), but may pursue different `variants' of this goal (e.g. collect apples vs collect oranges). Importantly, agents' goals may not be known a priori to one another. Our contributions are threefold. First, we offer an initial description and formalisation of this setting. Second, we extend two popular task environments used in AHT research to support heterogeneous goals. Finally, we propose \textbf{GRILL} (Goal selection by RL with Imitation for Low-Level control), a novel hierarchical method that achieves higher returns than baseline methods across these two environments, and demonstrates greater sensitivity to different levels of cooperative opportunity. 

\section{Related work}\label{sec:related-work}

\begin{figure*}[t]
    \centering
    \includegraphics[width=0.8\linewidth]{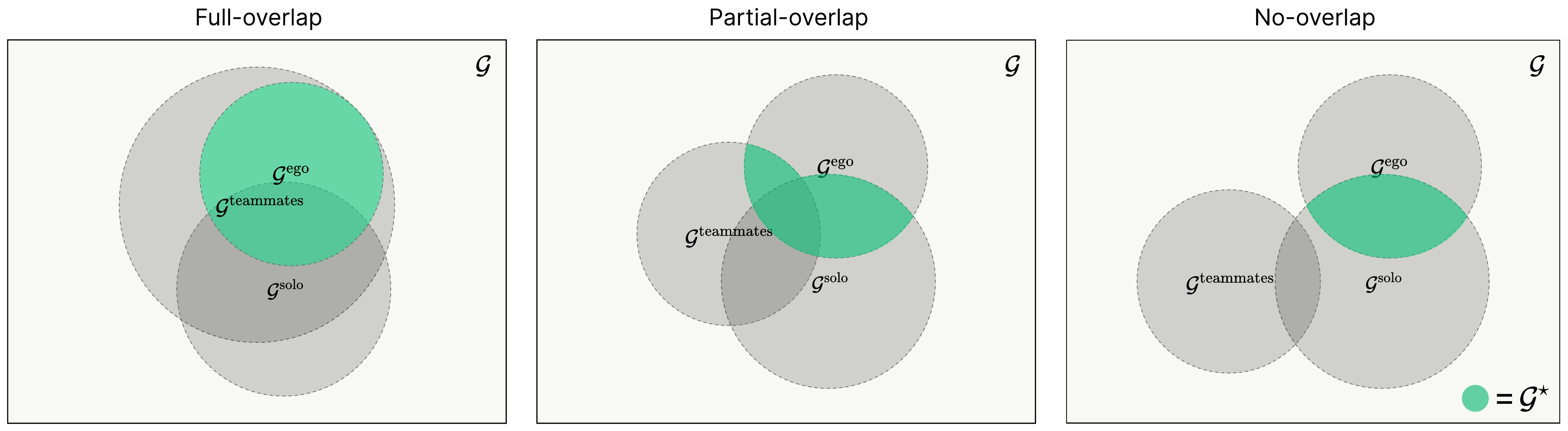}
    \caption{An illustration of the goal space under three different scenarios, with the set of `worthwhile' (rewarding and in-principle achievable) goals highlighted in green. $\mathcal{G}^\text{ego}$ denotes the set of goals that would produce reward for the ego agent; $\mathcal{G}^\text{teammates}$ the set of goals that would produce reward for at least one teammate; and $\mathcal{G}^\text{solo}$ the set of goals that can be achieved through the effort of a single agent.}
    \label{fig:goal-sets}
\end{figure*}

\subsection{Ad hoc teamwork}
The field of ad hoc teamwork (AHT) deals with the problem of developing agents capable of collaborating with previously unseen `teammates', in the absence of prior coordination or communication \cite{mirsky2022survey}. AHT shares some basic elements with the field of multi-agent reinforcement learning (MARL); but where MARL typically assumes control of all agents in the environment, in AHT we control only a single agent (often called the `AHT agent' or `learner'; we will use `ego agent' throughout). At its core, AHT is about adapting to some source of behavioural diversity across a population of different potential teammates. In practice, this has mostly taken the form of teammates having different `styles' of policy, while being otherwise homogeneous, controlled by either simple heuristics or pre-trained (frozen) RL policies. Many approaches have involved explicit inference and representation of teammate policy types; traditionally via forms of Bayesian belief-updating over a discrete teammate space \cite{gmytrasiewicz2005framework, barrett2011empirical, albrecht2013game, albrecht2016belief}, or more recently using neural-network-based encoders to learn latent policy representations \cite{rabinowitz2018machine, papoudakis2020variational, rahman2023general}. Some recent work has also considered heterogeneity in teammates' \emph{capabilities} (e.g. via different robot morphologies) \cite{liemhetcharat2014weighted, liu2024leveraging}. However, while the definition of the AHT problem allows for agents to have separate individual reward functions (within a broader cooperative task), the vast majority of existing work assumes a single reward function that is common to the ego agent and all their teammates. While this assumption accurately describes some real-world cooperative tasks, it is limiting---human environments typically feature people pursuing a range of different goals with varying degrees of overlap; an agent designed to operate in such environments should not blindly assume that everyone it encounters is a potential collaborator. In the current work, we focus on this underexplored angle of diversity by considering a setting in which agents can differ in the goals they pursue under a given high-level task. 

\subsection{Hierarchical reinforcement learning}\label{subsec:hrl}
Where traditional RL considers actions at only a single resolution, hierarchical methods operate over multiple levels of temporal abstraction. By breaking down the learning problem in this way, HRL algorithms can in principle offer better sample complexity and generalisation performance relative to their non-hierarchical counterparts. While recent research has focused on the more general `options' framework \cite{sutton1999between, bacon2017option, barreto2019option, klissarov2021flexible}, of more relevance to our work is the earlier Feudal RL (FRL) \cite{dayan1992feudal}. In FRL, the agent maintains distinct high-level and low-level policies, where the high-level policy is used to select `subgoals', and the low-level policy is used to select elementary actions conditioned on a given subgoal. FRL works well when the separation between these two levels is clearly defined; it is thus a sensible starting point for the current setting, where each environment is characterised by a fixed set of goals. The main difference with respect to the method we introduce here is that FRL optimises the low-level policy via a reward function of the form $R(s,a \mid s) = \mathbb{I}(s = g)$ (see the following section), while we use behavioural cloning with learned goal labels (using RL only to learn high-level goal selection).

\subsection{Goal-conditioned policy learning}
Another family of methods that extends the basic RL paradigm is goal-conditioned reinforcement learning (GCRL, sometimes also referred to as multi-goal or multi-objective RL), where the agent learns an action policy conditioned on different `goals' $g$; i.e. $\pi(a\mid o, g)$ \cite{kaelbling1993learning, liu2022goal}. The low-level policy in Feudal RL \cite{dayan1992feudal} (see previous section) is one early example of GCRL, but not all GCRL approaches involve a hierarchical structure. The goal space is typically operationalised as a subset of the state space (where the agent is rewarded only for reaching the particular state corresponding to their current goal)---but may also be represented in natural language \cite{chan2019augmenting, colas2020language}, or as levels of expected episode return \cite{chen2021decision, mccallum2023is}. While most GCRL approaches involve pre-specified goals, some recent work has also explored the possibility of agents learning to generate their own goals \cite{colas2022autotelic}.

Of more direct relevance to our own method is work that incorporates the idea of goal-conditioned policies into \emph{imitation} learning. For example, \citet{lynch2019learning} use a combination of self-supervised representation learning and supervised behaviour learning to obtain goal-conditioned robot policies from unstructured play data. \citet{ghosh2020learning} employ goal-conditioned behaviour cloning in a self-imitation paradigm, where the agent's previous trajectories are used in supervised learning as successful examples for the states they \emph{actually} reached (regardless of the states the agent `intended' to reach). Other works have explored goal-conditioned imitation learning with a version of the GAIL algorithm \cite{ding2020goal}, using diffusion policies \cite{reuss2023goal}, or transformer architectures \cite{sundaresan2025rt}. 

\section{Problem setting}\label{sec:problem-setting}

We consider the problem in which an \emph{ego agent} must act to achieve goals in an environment populated by other agents (we will use the term \emph{teammates} for consistency with the AHT literature) which have potentially mixed objectives. Note that while the experimental results we present use only a single teammate, for purposes of generality we lay out the arbitrary $n$-teammate case here.

Formally, we define our problem within the framework of Partially-Observable Stochastic Games (POSG). A POSG is defined by the tuple $\langle \mathcal{N}, \mathcal{S}, \mathcal{T}, \{\mathcal{A}^i\}, \{\mathcal{O}^i\}, \{\mathcal{Z}^i\}, \{r^i\}, \gamma\rangle$ where $\mathcal{N}$ is the set of agents (with $i \in \mathcal{N}$), $\mathcal{S}$ is the state space, and $\mathcal{T} : \mathcal{S} \times \vec{\mathcal{A}} \mapsto \Delta (\mathcal{S})$ is the transition function, with $\vec{\mathcal{A}} = \mathcal{A}_1 \times ... \times A_N$ denoting the joint action space. For each agent $i$ we have an action space $\mathcal{A}^i$, an observation space $\mathcal{O}^i$, an observation function $\mathcal{Z}^i : S \mapsto \mathcal{O}^i$ and a reward function $r^i : S \times \vec{\mathcal{A}} \mapsto \mathbb{R}$. $\gamma \in [0, 1]$ is the discount factor. 

In general, we consider an environment which offers a set $\mathcal{G}$ of different possible goals; some of which can be achieved by a single agent acting alone, and some of which require cooperation between two or more agents. Formally, we can say that a goal $g$ is defined by a tuple $(\Omega_g, r_g)$, where $\Omega_g \subseteq \mathcal{S} \times \mathcal{A} \times \mathcal{S}$ is a particular subset of the one-step transition space and $r_g$ is a scalar reward value. For example, in our foraging environment (see Section~\ref{sec:experiments}), the goal `collect apples' would correspond to $\Omega_g = \{(s,a,s') : \text{is\_adjacent\_apple}(s) \land a = \text{collect} \land \text{apple\_collected}(s') \}$. Each agent's objectives are then characterised by a binary mask over $\mathcal{G}$---i.e., for each goal $g \in \mathcal{G}$, an agent $i$ will receive a reward of either $r_g$ or 0 for executing a transition $\in \Omega_g$. We can therefore express all individual differences in reward in terms of the goal subsets $\mathcal{G}^i \subseteq \mathcal{G}$ for which different agents receive nonzero reward. For example, in our foraging example, we might have $\mathcal{G} = \{\text{collect apples}, \text{collect oranges}, \text{collect plums}\}$ and $\mathcal{G}^i = \{\text{collect apples}\}$. We also make the assumption that goals which require cooperation are always more rewarding than those that don't, i.e. $r_g > r_{g'} \ \forall \ g \notin \mathcal{G}^\text{solo}, g' \in \mathcal{G}^\text{solo}$---intended to reflect the reality that people can typically achieve more acting together than alone. Each agent in the environment expresses an observable cue $\phi^i$ that noisily signals their goals, i.e. $\phi^i \sim \mathcal{N}(\mathcal{G}^i, \sigma^2 I)$. Unless otherwise stated, we will use $\sigma^2 = .05$ throughout our experiments. 

\begin{figure*}[t]
    \centering
    \includegraphics[width=0.8\linewidth]{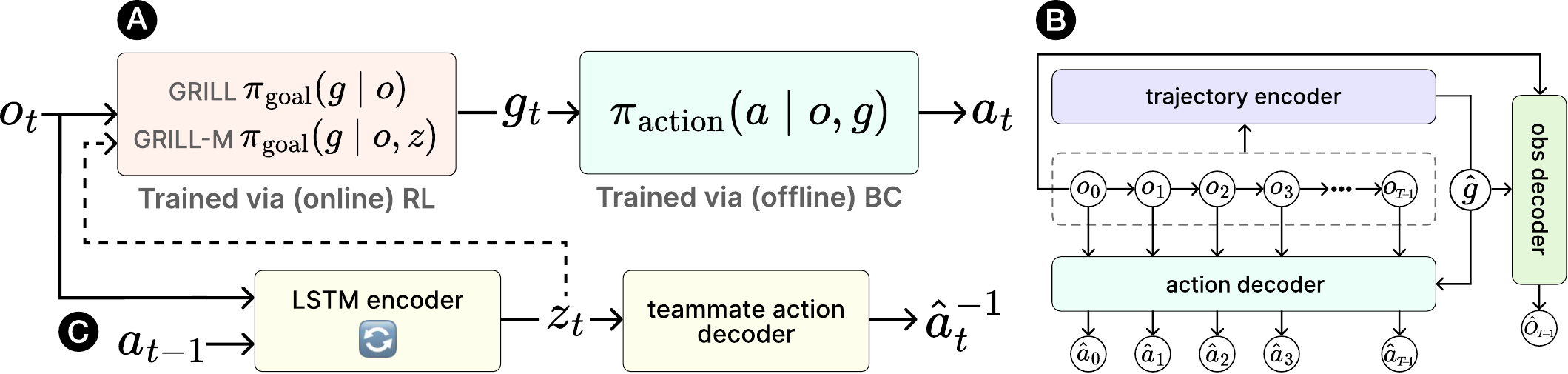}
    \caption{\textbf{(A)} The hierarchical architecture of GRILL \textbf{(B)} The encoder-decoder architecture optimised offline in stage 1, from which the action decoder becomes the low-level policy $\pi_\text{action}$ in stage 2 \textbf{(C)} The auxiliary modelling component used in GRILL-M (but not GRILL)}
    \label{fig:architecture}
\end{figure*}

We can think about rational behaviour in this setting through the lens of the goals that are \emph{worthwhile} for the ego agent to pursue (i.e. both rewarding and potentially achievable). The set of such goals can be written as $\mathcal{G}^\star = \mathcal{G}^\text{ego} \cap \big( \mathcal{G}^\text{teammates} \cup \mathcal{G}^\text{solo} \big)$, where $\mathcal{G}^\text{ego}$ is the ego agent's goal subset, $\mathcal{G}^\text{teammates}$ is the union of all teammates' goal subsets, and $\mathcal{G}^\text{solo}$ is the subset of goals that don't require cooperation. Figure~\ref{fig:goal-sets} illustrates how $\mathcal{G}^\star$ (highlighted in green) changes across the three basic scenarios that arise from this general setting:
\begin{enumerate}
    \item \textbf{full-overlap} (all of the ego agent's goals are shared by at least one teammate): $\mathcal{G}^\text{ego} \subseteq \mathcal{G}^\text{teammates}$
    \item \textbf{partial-overlap} (at least one but not all of the ego agent's goals are shared by at least one teammate): \\ $\mathcal{G}^\text{ego} \cap \mathcal{G}^\text{teammates} \neq \emptyset, \ \mathcal{G}^\text{ego} \not\subset \mathcal{G}^\text{teammates}$
    \item \textbf{no-overlap} (none of the ego agent's goals is shared by any teammate): $\mathcal{G}^\text{ego} \cap \mathcal{G}^\text{teammates} = \emptyset$
\end{enumerate}
To succeed at the overall task, the ego agent should be able to navigate \textbf{all} of these scenarios---determining when to pursue collaborative goals and when to act by themselves.

\section{Method}\label{sec:method}

We present two methods: GRILL and GRILL-M, a variant that incorporates a version of the auxiliary teammate modelling component from LIAM \cite{papoudakis2021agent}. The core idea behind GRILL is to separate the problems of \textbf{(1)} learning which goals to pursue given a particular environment state, and \textbf{(2)} learning which actions to take in order to achieve those goals. That is, rather than learning a single end-to-end policy, we learn separately a `high-level' policy for goal selection, and a `low-level' policy for goal-conditioned action selection. This kind of hierarchical structure is not new (see Section~\ref{subsec:hrl})---rather, our insight is that the optimal low-level policy is universal to all agents in the population, whereas the optimal high-level policy depends on the goals of both the ego agent and their current teammate. We can therefore use a two-stage process that combines imitation and reinforcement learning (illustrated in Figure~\ref{fig:architecture}):

\textbf{Stage 1}: We first collect a small offline dataset $\mathcal{D} = \{o_t, a_t\}$ of observations and actions from randomly sampled heuristic agents. $\mathcal{D}$ is then split into a set of fixed-length trajectories $\{\tau\}$, where each trajectory corresponds to the pursuit of a single goal; these are then used to optimise an encoder-decoder model. For each trajectory, the encoder produces a discrete goal label. Given the discrete encoding, one decoder then tries to predict the teammate's action at each point along the trajectory from the preceding observation; a second decoder tries to predict the final observation from the first. The whole system is optimised solely to reconstruct actions and observations from $\mathcal{D}$, with no explicit goal information provided:
\begin{align}\label{eq:bc-objective}
    \mathcal{L}_a = -\sum_{t=0}^{T-1} \log\text{dec1}_{\theta_\text{dec1}} \big(a_t \mid o_t, \hat{g} = \text{enc}_{\theta_\text{enc}}(\tau)\big) \nonumber \\
    \mathcal{L}_o = \Big\Vert o_{T-1} - \text{dec2}_{\theta_\text{dec2}}\big(o_0, \hat{g} = \text{enc}_{\theta_\text{enc}}(\tau)\big)\Big\Vert_2^2 \nonumber \\
    \min_{\theta_\text{enc}, \theta_\text{dec1}, \theta_\text{dec2}} \expect_{\tau \sim \mathcal{D}}\Big[ \lambda \mathcal{L}_a + (1-\lambda)\mathcal{L}_o\Big]
\end{align}
where $\lambda$ is a scalar that trades off between the two prediction objectives. After training, we discard the encoder and the observation decoder while retaining the action decoder as our low-level goal-conditioned policy $\pi_\text{action}$. 

\textbf{Stage 2}: we use PPO to learn a high-level policy $\pi_\text{goal}$ mapping the current observation to a discrete goal, the output of which is used to condition the BC policy from the previous stage. For GRILL-M, this stage also involves an auxiliary objective, where the ego agent is trained to predict their teammate's actions from their own observations and actions via an LSTM encoder-decoder with objective
\begin{equation}
    \min_{\theta_\text{enc}, \theta_\text{dec}} -\text{log} \ \text{dec}_{\theta_\text{dec}}\big(a^{-1}_t|z_t = \text{enc}_{\theta_\text{enc}}(o_t, a_{t-1})\big)
\end{equation}
where $\pi_\text{goal}$ is learned over the augmented space $\mathcal{O} \times \mathcal{Z}$. This is almost identical to the modelling component of LIAM \cite{papoudakis2021agent}, with the sole difference that since our environments are fully observable, we remove the additional prediction of teammate observations.

\section{Experiments}\label{sec:experiments}

\subsection{Environments}
\begin{figure}[t]
    \centering
    \includegraphics[width=0.8\linewidth]{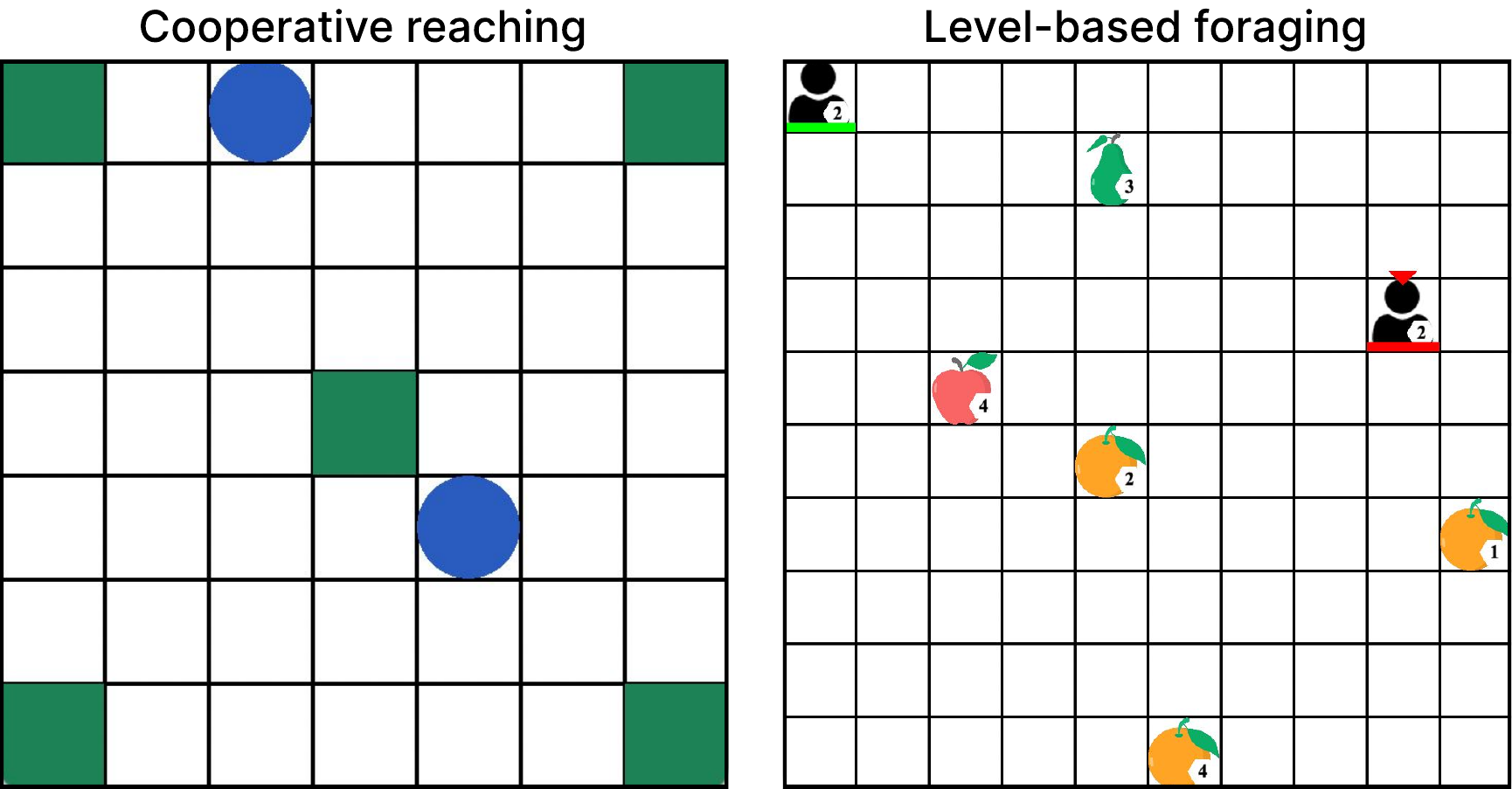}
    \caption{Example frames from the two AHT environments we extend.}
    \label{fig:environments}
\end{figure}

We extend two commonly used, fully observable AHT environments to incorporate the notion of goal heterogeneity. Example states of both are shown in Figure~\ref{fig:environments}.

\textbf{Cooperative reaching:} Cooperative reaching is a simple gridworld task where two agents must navigate to and jointly occupy one of four reward-producing corner tiles \cite{rahman2023generating, rahman2024minimum}. In the original version, different corners are associated with different amounts of reward, but the reward for reaching a given corner does not vary between agents. In our version, each corner tile yields a reward of either 1 or 0 for different agents; we also add an additional goal tile in the centre of the grid which can be successfully reached by a single agent (i.e. is in $\mathcal{G}^\text{solo}$) but produces a lower reward of 0.2. All goal tiles act as absorbing states---once an agent has reached one of these five tiles, they are unable to take any more actions. The ego agent receives observations containing their own location, and the location and observable cue ($\phi^i$) of their teammate. The action space consists simply of the four cardinal directions plus a no-op action. 

\textbf{Level-based foraging:} Level-based foraging (LBF) is another gridworld environment, where agents cooperate to collect items. Each agent and item have an associated `level', where an item with level $l$ can only be collected by the joint efforts of a group of agents whose levels sum to $\geq l$ \cite{albrecht2013game, yourdshahi2018towards, rahman2023general}. Our version of the environment extends this: rather than a single item type, we have three different `fruits' (which can also still vary in level). The set of possible goals that the environment can contain is therefore given by $\mathcal{G} = \{(\text{type}, l) \ \forall \ \text{type} \in \{\text{apple}, \text{orange}, \text{plum}\}, l \in [1, l_\text{max}]\}$ where $l_\text{max}$ is the maximum item level. The ego agent receives observations containing the location, type and level of all fruits in the environment, as well as their own location and level, plus the location, level and observable cue ($\phi^i$) of their teammate. The action space consists of the four cardinal directions, a `collect' action (to attempt fruit collection) and a no-op action. For simplicity, our experiments fix both ego agent and teammate at level 1. 

\subsection{Baselines}
We evaluate GRILL against three baselines: PPO, LIAM and OMG. PPO \cite{schulman2017proximal} is a general RL algorithm that for our purposes serves as a floor for what can be achieved by a method not tailored in any way to cooperative (or even multi-agent) settings. LIAM \cite{papoudakis2021agent} and OMG \cite{yu2024opponent} are two recent methods that use some form of agent modelling to achieve competitive performance on AHT tasks similar to those we use here. Both LIAM and OMG train the ego agent's policy over an `augmented' observation space incorporating a latent teammate representation produced by an auxiliary network. In LIAM, this is produced by a recurrent encoder-decoder trained to predict teammates' current observations and actions from the ego agent's own (observation, action) pairs. OMG instead uses the latent representation from a conditional VAE optimised to model the teammate's `subgoal' (as a feature embedding of some future state). We note that while the LIAM authors used A2C, and OMG used different RL algorithms for different tasks, our implementations of both use PPO as a backbone for consistency across methods. For each environment, we also compare against an `oracle' policy which uses a shortest-path heuristic guided by full knowledge of all agents' goals and the base rewards. All methods (aside from the oracle) were trained for $10^7$ timesteps for cooperative reaching, and $5\times 10^7$ timesteps for level-based foraging. 

\begin{figure*}[t]
    \centering
    \includegraphics[width=0.8\linewidth]{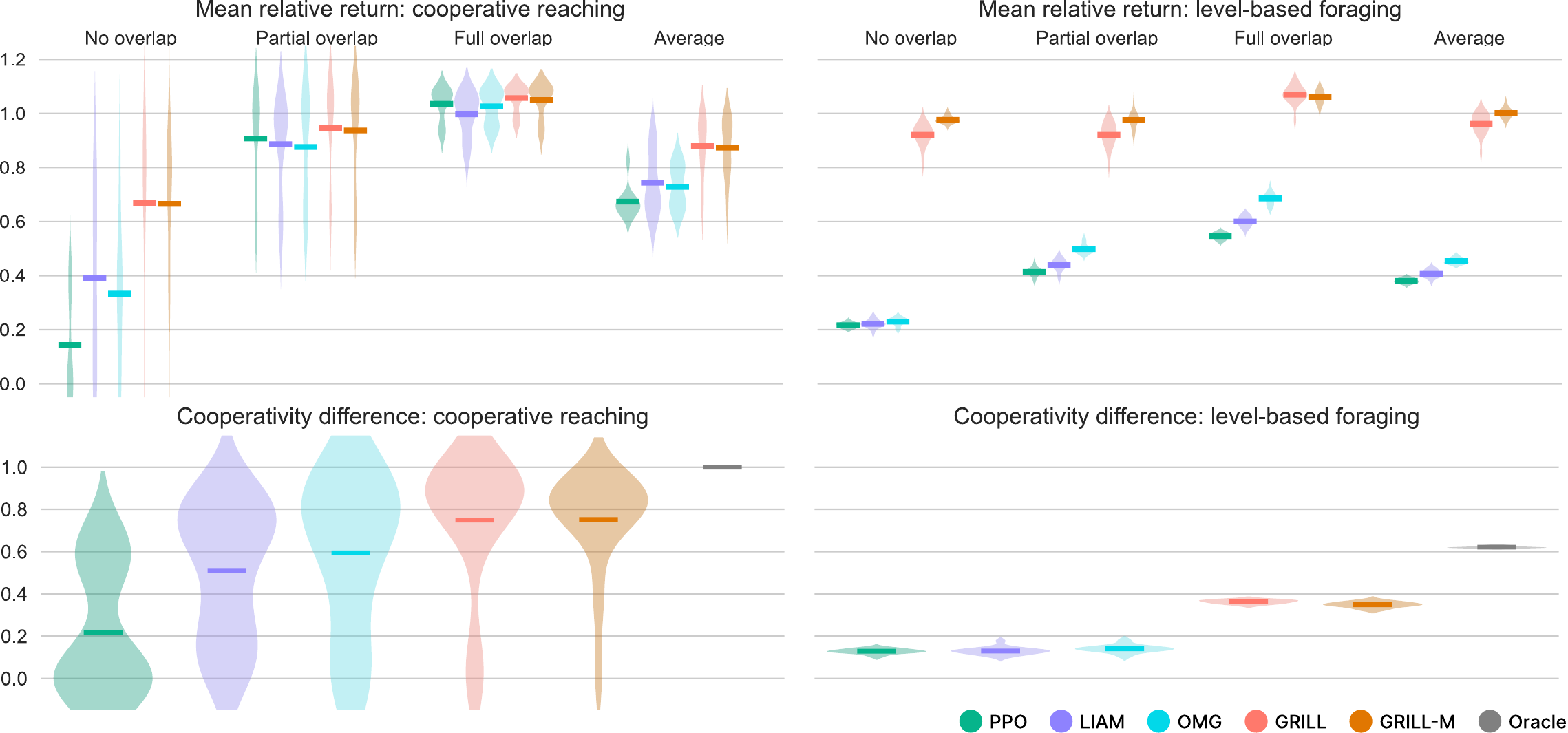}
    \caption{\textbf{Top:} evaluation returns relative to oracle policy, measured over 1000 episodes $\times$ the 3 scenarios $\times$ 20 independent training runs. Mean values are given by the horizontal markers, while underlaid violin plots show the KDE. Agents were trained for 1e7 and 5e7 timesteps on CR and LBF respectively. \textbf{Bottom:} from the same set of evaluation episodes, the difference in proportion of non-solo goals pursued between the full- and no-overlap scenarios.}
    \label{fig:returns-and-cd}
\end{figure*}

\subsection{Evaluation}\label{subsec:eval}
In addition to the standard evaluation metric of average episode return, we also want to take a deeper look at how trained policies deal with the three scenarios enumerated in Section~\ref{sec:problem-setting}. To do this, we look at which goals the ego agent \emph{attempts} to reach during evaluation episodes---i.e. which absorbing tiles does it navigate to in CR, and which fruits does it try to collect in LBF. We consider three distinct failure modes relating to goal selection. The first (and most basic) is to seek goals outside of $\mathcal{G}^\text{ego}$; i.e., goals which won't even yield any reward if achieved. The second is to be \emph{over-collaborative} by seeking goals that are in $\mathcal{G}^\text{ego}$ but unachievable ($\notin \mathcal{G}^\text{solo} \cup \mathcal{G}^\text{teammates}$; note that this failure mode doesn't exist in the full-overlap scenario). The third and most subtle is to be \emph{under-collaborative} by not seeking achievable cooperative goals when such goals exist (note that this failure mode doesn't exist in the no-overlap scenario). A successful agent should avoid all three of these failure modes.

We measure the distribution of ego agent goal choices over four distinct subsets that allow us to distinguish these three failure modes (see Figures~\ref{fig:goal-sets} and \ref{fig:goal-dists}). As an additional summary metric, we also report the `cooperativity difference' ($\Delta_\text{coop}$) as the difference in the proportion of pursued goals $\notin \mathcal{G}^\text{solo}$ between the full- and no-overlap scenarios---where a higher value indicates a greater flexibility in the ego agent's strategy.

\section{Results}
\subsection{GRILL achieves higher returns than all baselines}

Figure~\ref{fig:returns-and-cd} (top row) shows the average return after training, relative to each environment's oracle policy, for the three scenarios enumerated in Section~\ref{sec:problem-setting}. Overall, we find that GRILL and GRILL-M outperform all baselines across every scenario in both environments. 

Taking a closer look at the results for cooperative reaching (left), we can see that for partial- and full-overlap all methods do similarly well, achieving returns close to or even slightly above the oracle policy. In the no-overlap scenario, all methods (including ours) do considerably worse---interestingly, this is the case in which we see the largest difference between PPO and the other two baselines. Across all scenarios in cooperative reaching, we find no discernable difference between GRILL and GRILL-M. On the harder environment of LBF (right), we observe a striking gap in performance between GRILL and the three baselines across all scenarios. Of the baselines, OMG consistently outperforms LIAM and PPO on LBF, although not by a huge margin. Interestingly, here we do see a difference between the two variants of our method, with GRILL-M eking out a small advantage in the no and partial overlap scenarios. 

\subsection{GRILL pursues more worthwhile goals}
\begin{figure*}[t]
    \centering
    \includegraphics[width=0.8\linewidth]{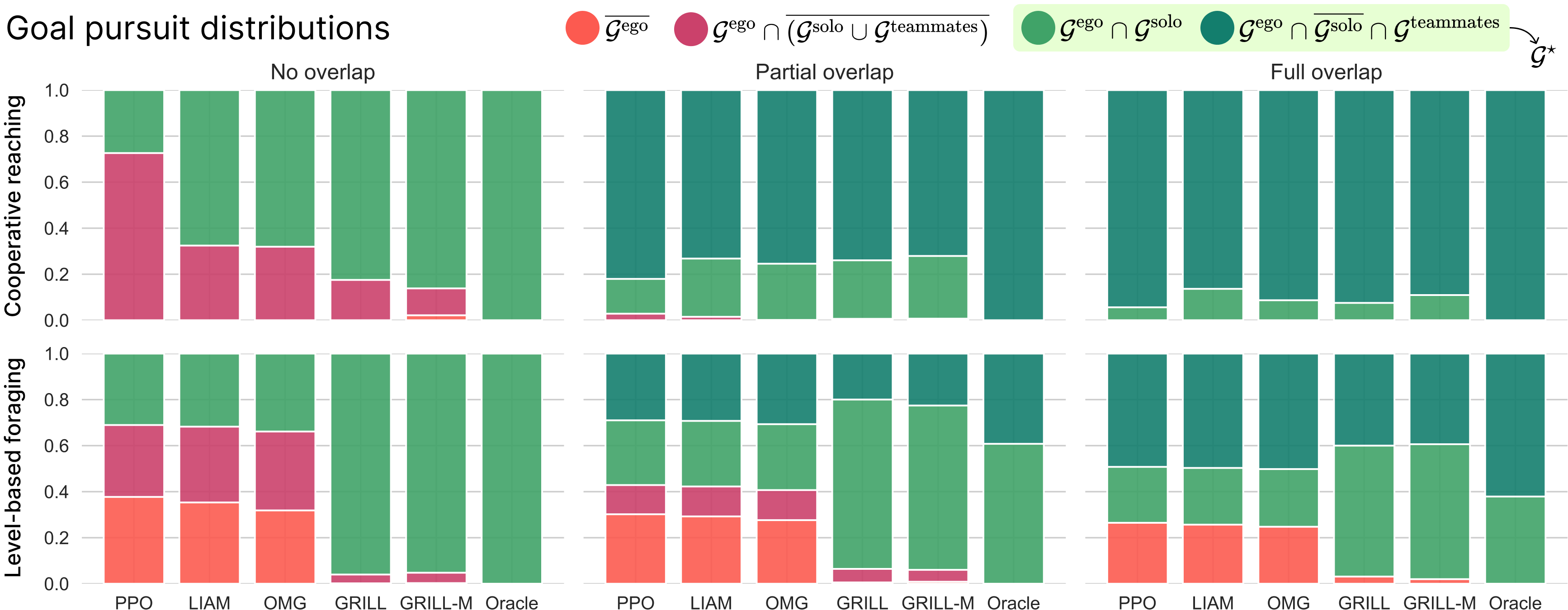}
    \caption{The distribution of goals attempted by the ego agent during evaluation. For CR, `attempt' means the agent occupied one of the goal tiles; for LBF, it means the agent used the `collect' action while adjacent to a fruit.}
    \label{fig:goal-dists}
\end{figure*}

\begin{figure}
    \centering
    \includegraphics[width=0.8\linewidth]{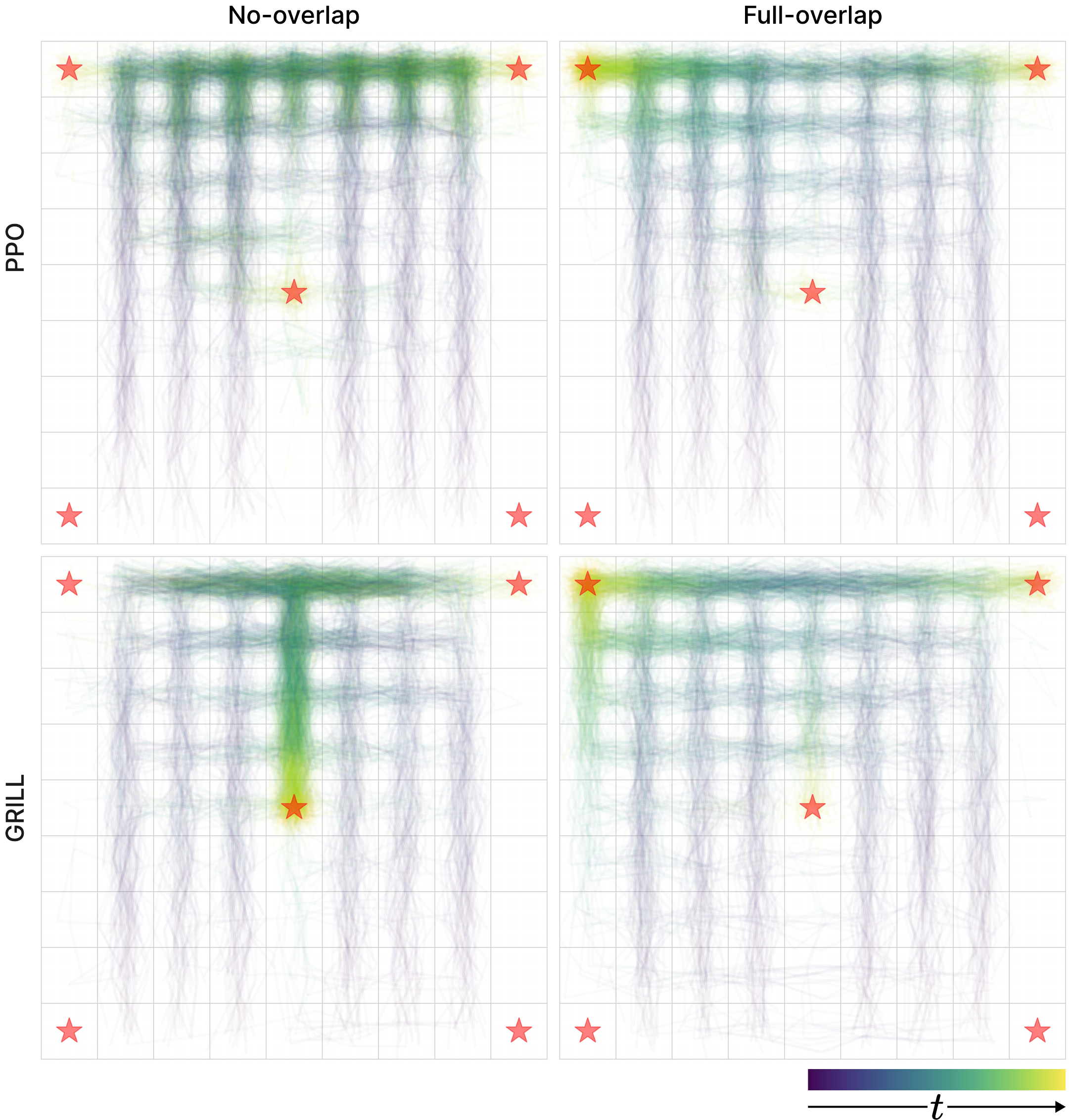}
    \caption{A visualisation of trajectories sampled from trained PPO and GRILL agents in the no- and full-overlap scenarios of the cooperative reaching environment. Trajectories are sampled from random initial locations, and a small amount of noise is applied to each path in order to better visualise the distributions.}
    \label{fig:trajectories}
\end{figure}

Figure~\ref{fig:goal-dists} shows, for both environments, the distribution of goals pursued by the trained ego agent during evaluation---where we have partitioned the goal space into four subsets designed to illuminate the three failure modes outlined in Section~\ref{subsec:eval}. In cooperative reaching, all agents were able to avoid the first failure mode, seeking only goals that would be rewarding if achieved. In the no-overlap scenario, all agents to varying degrees fell victim to the second failure mode, pursuing cooperative goals even though it was futile to do so. This is especially pronounced for the PPO agent, which chose the correct tile less than 30\% of the time. In the partial- and full-overlap scenarios, the goal distributions were very similar across all four methods, with the agent being \emph{under}-cooperative relative to the optimal policy. Turning to LBF, we find that PPO, LIAM and OMG  exhibited the first failure mode across all three scenarios, in addition to the second failure mode in no- and partial-overlap, and the third failure mode in partial- and full-overlap. By contrast, GRILL avoids the first failure mode entirely, and the second almost entirely, seeking worthwhile goals over 90\% of the time across all three scenarios. However, it does still suffer from being under-cooperative in the partial and full overlap settings; and on this particular measure is actually slightly worse than the three baselines. Overall, agents trained using our method(s) selected a higher proportion of worthwhile goals than any baseline method, across both environments. For an additional visualisation of this result, Figure~\ref{fig:trajectories} compares trajectories from PPO and GRILL in cooperative reaching; highlighting a much greater difference in where the latter agent tends to end up between the no- and full-overlap scenarios. 

The bottom row of Figure~\ref{fig:returns-and-cd} shows the `cooperativity difference' $\Delta_\text{coop}$ for each method, i.e. the difference between how much the ego agent pursues cooperative goals between the full- and no-overlap scenarios. Consistent with the performance results, we find a higher $\Delta_\text{coop}$ value for GRILL than the baselines across both environments, and a higher $\Delta_\text{coop}$ for LIAM and OMG than PPO only for cooperative reaching.

\subsection{GRILL-M outperforms GRILL when teammate goal information is noisier}

\begin{figure*}
    \centering
    \includegraphics[width=0.8\linewidth]{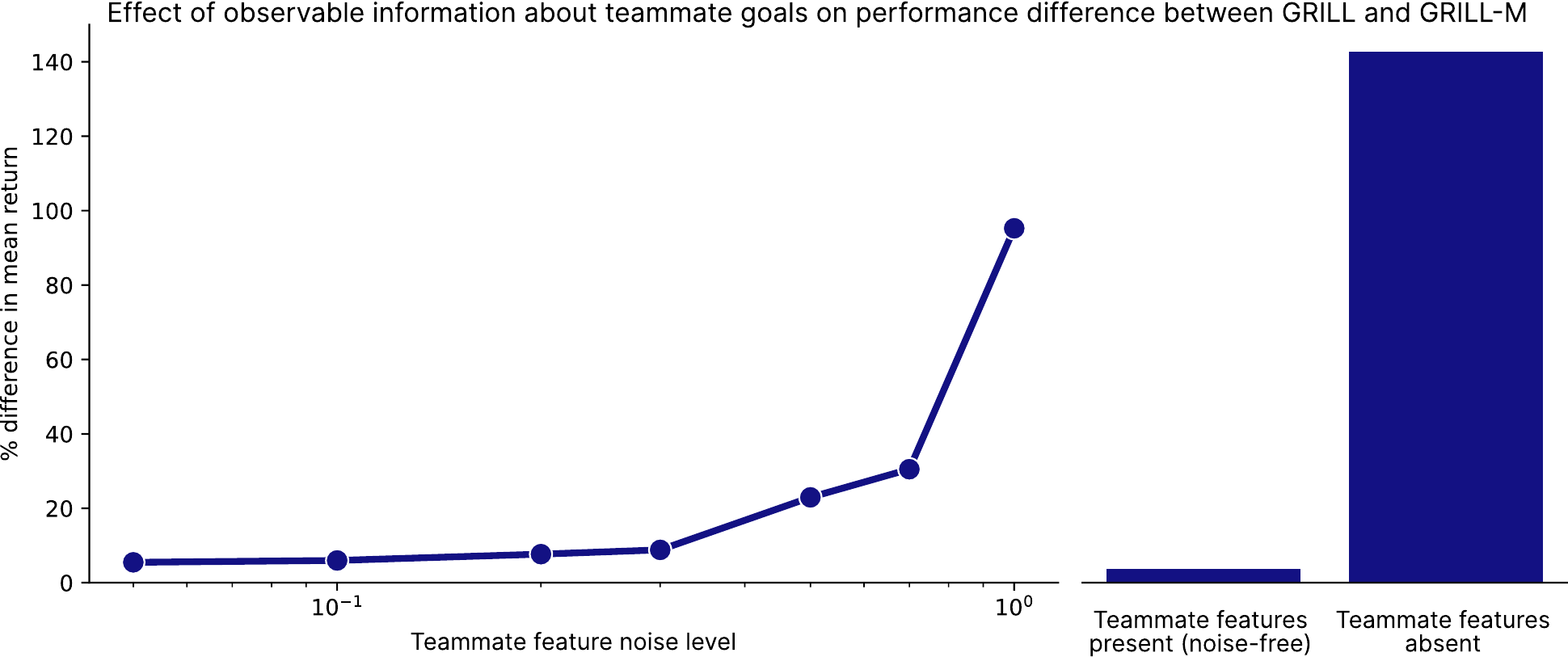}
    \caption{\textbf{Left:} the percentage increase in mean evaluation return in the LBF environment for GRILL-M vs GRILL, as a function of the amount of noise in the ego agent's observation of their teammate's goal vector. Returns are averaged over 1000 episodes $\times$ the 3 scenarios $\times$ 20 independent training runs. \textbf{Right:} the same increase compared for the extreme cases of no noise and teammate goal information entirely absent.}
    \label{fig:noise-results}
\end{figure*}

Finally, we ran a small additional experiment to try and understand the pattern of results observed between the two variants of our method. Our hypothesis was that in cooperative reaching, having access to information about teammate goals in the observation would be less important to performance, since the teammate's goals are more clearly evidenced by their behaviour. In LBF, where the larger space of possible behaviours renders the relationship between goals and actions less trivial, having this information in the observation should be more useful. If the latent representations capture information about teammates' goals, then they should aid performance to the extent that that information is not redundant with the `base' observation. 

To test this, we ran training and evaluation for GRILL and GRILL-M on the LBF environment, increasing the level of noise in the teammate's observable feature $\phi$ from 0.05 (the value used in our previous runs) up to 1.0. We also ran a more extreme comparison where we removed $\phi$ from the observation entirely. Figure ~\ref{fig:noise-results} shows that as the noise increases, the performance gap widens monotonically---with GRILL-M achieving 95.2\% higher average return at $\sigma^2 = 1.0$, up from only 5.5\% at $\sigma^2 = 0.05$. This increases further to $142.6\%$ when $\phi$ is removed entirely from the observation space. These results support our hypothesis about when the auxiliary latent representations are beneficial.

\section{Discussion}
In this paper, we have introduced a variant of the standard ad hoc teamwork problem in which agents can have heterogeneous goals that may or may not overlap in any given episode. Our motivation for studying this setting is to capture the intuition that in many real-world settings---where not all agents have common goals---people achieve better outcomes by understanding when collaboration is likely to prove fruitful. This is in contrast to most work in the AHT setting, which typically deals with the problem of adapting to teammates who might behave differently in pursuit of the same goal. 

As a secondary contribution, we presented GRILL, a novel hierarchical method for learning flexible policies in the heterogeneous-goals setting that works by combining imitation at the action-selection level with RL at the goal-selection level. Using extended versions of two popular AHT environments---cooperative reaching and level-based foraging---we showed that GRILL outperforms baseline methods by virtue of more flexible goal selection. In a small additional experiment, we found that the effect on performance of an auxiliary prediction objective (a la LIAM) depends on the availability of teammate goal information in the observation space. 

While our focus here is on capturing a particular dimension of flexible cooperation, we feel that the core idea behind GRILL should be widely applicable to any multi-goal, multi-agent setting, regardless of whether or not it involves collaboration. For example, some explicitly competitive domains may present an analogous decision-making problem where an agent must choose whether to pursue a high-value goal for which there is a lot of competition, or settle for a lower-value but less contested target. We plan to test this idea in future work, as well as explore connections to human behaviour learning. Follow-up research may also address limitations of the current work---such as the fact that our experiments were limited to fully-observable environments with only two agents, or that the low-level policy was learned offline in a prior phase rather than online alongside the high-level policy. 

\bibliography{aaai2026}

@article{tomasello2012two,
  title={Two key steps in the evolution of human cooperation: The interdependence hypothesis},
  author={Tomasello, Michael and Melis, Alicia P and Tennie, Claudio and Wyman, Emily and Herrmann, Esther},
  journal={Current Anthropology},
  volume={53},
  number={6},
  pages={673--692},
  year={2012},
  publisher={University of Chicago Press Chicago, IL}
}

@book{henrich2018secret,
  author = {Henrich, Joseph},
  title = {The Secret of Our Success: How Culture Is Driving Human Evolution, Domesticating Our Species, and Making Us Smarter},
  publisher = {Princeton University Press},
  year = {2018},
  doi = {10.1515/9781400873296}
}

@article{sutton1999between,
	title = {Between {MDPs} and semi-{MDPs}: {A} framework for temporal abstraction in reinforcement learning},
	volume = {112},
	issn = {0004-3702},
	number = {1},
	journal = {Artificial Intelligence},
	author = {Sutton, Richard S. and Precup, Doina and Singh, Satinder},
	year = {1999},
	pages = {181--211},
}

@inproceedings{bacon2017option,
    author = {Bacon, Pierre-Luc and Harb, Jean and Precup, Doina},
    title = {The option-critic architecture},
    year = {2017},
    publisher = {AAAI Press},
    booktitle = {Proceedings of the Thirty-First AAAI Conference on Artificial Intelligence},
    pages = {1726–1734},
    series = {AAAI'17}
}

@article{barreto2019option,
	title = {The option keyboard: {Combining} skills in reinforcement learning},
	volume = {32},
	journal = {Advances in Neural Information Processing Systems},
	author = {Barreto, André and Borsa, Diana and Hou, Shaobo and Comanici, Gheorghe and Aygün, Eser and Hamel, Philippe and Toyama, Daniel and Mourad, Shibl and Silver, David and Precup, Doina},
	year = {2019},
}

@inproceedings{klissarov2021flexible,
	title = {Flexible {Option} {Learning}},
	volume = {34},
	booktitle = {Advances in {Neural} {Information} {Processing} {Systems}},
	author = {Klissarov, Martin and Precup, Doina},
	year = {2021},
	pages = {4632--4646},
}

@inproceedings{dayan1992feudal,
     author = {Dayan, Peter and Hinton, Geoffrey E},
     booktitle = {Advances in Neural Information Processing Systems},
     editor = {S. Hanson and J. Cowan and C. Giles},
     pages = {},
     publisher = {Morgan-Kaufmann},
     title = {Feudal Reinforcement Learning},
     volume = {5},
     year = {1992}
}

@inproceedings{liu2022goal,
  title     = {Goal-Conditioned Reinforcement Learning: Problems and Solutions},
  author    = {Liu, Minghuan and Zhu, Menghui and Zhang, Weinan},
  booktitle = {Proceedings of the Thirty-First International Joint Conference on
               Artificial Intelligence, {IJCAI-22}},
  publisher = {International Joint Conferences on Artificial Intelligence Organization},
  editor    = {Lud De Raedt},
  pages     = {5502--5511},
  year      = {2022},
  month     = {7},
}

@article{colas2022autotelic,
	title = {Autotelic {Agents} with {Intrinsically} {Motivated} {Goal}-{Conditioned} {Reinforcement} {Learning}: {A} {Short} {Survey}},
	volume = {74},
	journal = {J. Artif. Int. Res.},
	author = {Colas, Cédric and Karch, Tristan and Sigaud, Olivier and Oudeyer, Pierre-Yves},
	year = {2022},
}

@inproceedings{kaelbling1993learning,
  title={Learning to Achieve Goals},
  author={Leslie Pack Kaelbling},
  booktitle={International Joint Conference on Artificial Intelligence},
  year={1993},
  url={https://api.semanticscholar.org/CorpusID:5538688}
}

@inproceedings{sundaresan2025rt,
	title = {{RT}-{Sketch}: {Goal}-{Conditioned} {Imitation} {Learning} from {Hand}-{Drawn} {Sketches}},
	shorttitle = {{RT}-{Sketch}},
	booktitle = {Proceedings of {The} 8th {Conference} on {Robot} {Learning}},
	publisher = {PMLR},
	author = {Sundaresan, Priya and Vuong, Quan and Gu, Jiayuan and Xu, Peng and Xiao, Ted and Kirmani, Sean and Yu, Tianhe and Stark, Michael and Jain, Ajinkya and Hausman, Karol and Sadigh, Dorsa and Bohg, Jeannette and Schaal, Stefan},
	year = {2025},
	pages = {70--96},
}

@misc{ding2020goal,
	title = {Goal-conditioned {Imitation} {Learning}},
	publisher = {arXiv},
	author = {Ding, Yiming and Florensa, Carlos and Phielipp, Mariano and Abbeel, Pieter},
	year = {2020},
	note = {arXiv:1906.05838 [cs]},
}

@misc{reuss2023goal,
	title = {Goal-{Conditioned} {Imitation} {Learning} using {Score}-based {Diffusion} {Policies}},
	publisher = {arXiv},
	author = {Reuss, Moritz and Li, Maximilian and Jia, Xiaogang and Lioutikov, Rudolf},
	year = {2023},
	note = {arXiv:2304.02532 [cs]},
}

@misc{lynch2019learning,
	title = {Learning {Latent} {Plans} from {Play}},
	publisher = {arXiv},
	author = {Lynch, Corey and Khansari, Mohi and Xiao, Ted and Kumar, Vikash and Tompson, Jonathan and Levine, Sergey and Sermanet, Pierre},
	year = {2019},
	note = {arXiv:1903.01973 [cs]},
}

@misc{ghosh2020learning,
	title = {Learning to {Reach} {Goals} via {Iterated} {Supervised} {Learning}},
	url = {http://arxiv.org/abs/1912.06088},
	publisher = {arXiv},
	author = {Ghosh, Dibya and Gupta, Abhishek and Reddy, Ashwin and Fu, Justin and Devin, Coline and Eysenbach, Benjamin and Levine, Sergey},
	year = {2020},
	note = {arXiv:1912.06088 [cs]},
}

@misc{mirsky2022survey,
	title = {A {Survey} of {Ad} {Hoc} {Teamwork} {Research}},
	publisher = {arXiv},
	author = {Mirsky, Reuth and Carlucho, Ignacio and Rahman, Arrasy and Fosong, Elliot and Macke, William and Sridharan, Mohan and Stone, Peter and Albrecht, Stefano V.},
	month = aug,
	year = {2022},
	note = {arXiv:2202.10450 [cs]},
}

@article{gmytrasiewicz2005framework,
    author = {Gmytrasiewicz, Piotr J. and Doshi, Prashant},
    title = {A framework for sequential planning in multi-agent settings},
    year = {2005},
    publisher = {AI Access Foundation},
    volume = {24},
    number = {1},
    journal = {J. Artif. Int. Res.},
    month = jul,
    pages = {49–79},
}

@inproceedings{barrett2011empirical,
    author = {Barrett, Samuel and Stone, Peter and Kraus, Sarit},
    title = {Empirical evaluation of ad hoc teamwork in the pursuit domain},
    year = {2011},
    publisher = {International Foundation for Autonomous Agents and Multiagent Systems},
    booktitle = {The 10th International Conference on Autonomous Agents and Multiagent Systems - Volume 2},
    pages = {567–574},
    series = {AAMAS '11}
}

@inproceedings{albrecht2013game,
    author = {Albrecht, Stefano V. and Ramamoorthy, Subramanian},
    title = {A game-theoretic model and best-response learning method for ad hoc coordination in multiagent systems},
    year = {2013},
    booktitle = {Proceedings of the 2013 International Conference on Autonomous Agents and Multi-Agent Systems},
    pages = {1155–1156},
    series = {AAMAS '13}
}

@article{albrecht2016belief,
    title = {Belief and truth in hypothesised behaviours},
    journal = {Artificial Intelligence},
    volume = {235},
    pages = {63-94},
    year = {2016},
    issn = {0004-3702},
    doi = {https://doi.org/10.1016/j.artint.2016.02.004},
    author = {Stefano V. Albrecht and Jacob W. Crandall and Subramanian Ramamoorthy},
}

@inproceedings{rabinowitz2018machine,
  title = 	 {Machine Theory of Mind},
  author =       {Rabinowitz, Neil and Perbet, Frank and Song, Francis and Zhang, Chiyuan and Eslami, S. M. Ali and Botvinick, Matthew},
  booktitle = 	 {Proceedings of the 35th International Conference on Machine Learning},
  pages = 	 {4218--4227},
  year = 	 {2018},
  editor = 	 {Dy, Jennifer and Krause, Andreas},
  volume = 	 {80},
  series = 	 {Proceedings of Machine Learning Research},
  month = 	 {10--15 Jul},
  publisher =    {PMLR},
}

@article{rahman2023general,
  author  = {Arrasy Rahman and Ignacio Carlucho and Niklas HÃ¶pner and Stefano V. Albrecht},
  title   = {A General Learning Framework for Open Ad Hoc Teamwork Using Graph-based Policy Learning},
  journal = {Journal of Machine Learning Research},
  year    = {2023},
  volume  = {24},
  number  = {298},
  pages   = {1--74},
}

@article{papoudakis2020variational,
  title={Variational autoencoders for opponent modeling in multi-agent systems},
  author={Papoudakis, Georgios and Albrecht, Stefano V},
  journal={arXiv preprint arXiv:2001.10829},
  year={2020}
}

@inproceedings{papoudakis2021agent,
    author = {Papoudakis, Georgios and Christianos, Filippos and Albrecht, Stefano V.},
    title = {Agent modelling under partial observability for deep reinforcement learning},
    year = {2021},
    booktitle = {Proceedings of the 35th International Conference on Neural Information Processing Systems},
    series = {NIPS '21}
}

@misc{liu2024leveraging,
      title={Leveraging Large Language Model for Heterogeneous Ad Hoc Teamwork Collaboration}, 
      author={Xinzhu Liu and Peiyan Li and Wenju Yang and Di Guo and Huaping Liu},
      year={2024},
      eprint={2406.12224},
      archivePrefix={arXiv},
      primaryClass={cs.RO},
      url={https://arxiv.org/abs/2406.12224}, 
}

@article{liemhetcharat2014weighted,
    title = {Weighted synergy graphs for effective team formation with heterogeneous ad hoc agents},
    journal = {Artificial Intelligence},
    volume = {208},
    pages = {41-65},
    year = {2014},
    author = {Somchaya Liemhetcharat and Manuela Veloso},
}

@misc{schulman2017proximal,
      title={Proximal Policy Optimization Algorithms}, 
      author={John Schulman and Filip Wolski and Prafulla Dhariwal and Alec Radford and Oleg Klimov},
      year={2017},
      eprint={1707.06347},
      archivePrefix={arXiv},
      primaryClass={cs.LG},
      url={https://arxiv.org/abs/1707.06347}, 
}

@inproceedings{yu2024opponent,
 author = {Yu, Xiaopeng and Jiang, Jiechuan and Lu, Zongqing},
 booktitle = {Advances in Neural Information Processing Systems},
 editor = {A. Globerson and L. Mackey and D. Belgrave and A. Fan and U. Paquet and J. Tomczak and C. Zhang},
 pages = {60531--60555},
 title = {Opponent Modeling based on Subgoal Inference},
 volume = {37},
 year = {2024}
}

@inproceedings{chen2021decision,
  title={Decision Transformer: Reinforcement Learning via Sequence Modeling},
  author={Lili Chen and Kevin Lu and Aravind Rajeswaran and Kimin Lee and Aditya Grover and Michael Laskin and P. Abbeel and A. Srinivas and Igor Mordatch},
  booktitle={Neural Information Processing Systems},
  year={2021},
  url={https://api.semanticscholar.org/CorpusID:235294299}
}

@misc{chan2019augmenting,
      title={ACTRCE: Augmenting Experience via Teacher's Advice For Multi-Goal Reinforcement Learning}, 
      author={Harris Chan and Yuhuai Wu and Jamie Kiros and Sanja Fidler and Jimmy Ba},
      year={2019},
      eprint={1902.04546},
      archivePrefix={arXiv},
      primaryClass={cs.LG},
}

@inproceedings{colas2020language,
 author = {Colas, C\'{e}dric and Karch, Tristan and Lair, Nicolas and Dussoux, Jean-Michel and Moulin-Frier, Cl\'{e}ment and Dominey, Peter and Oudeyer, Pierre-Yves},
 booktitle = {Advances in Neural Information Processing Systems},
 editor = {H. Larochelle and M. Ranzato and R. Hadsell and M.F. Balcan and H. Lin},
 pages = {3761--3774},
 title = {Language as a Cognitive Tool to Imagine Goals in Curiosity Driven Exploration},
 volume = {33},
 year = {2020}
}

@inproceedings{yourdshahi2018towards,
	title = {Towards {Large} {Scale} {Ad}-hoc {Teamwork}},
	booktitle = {2018 {IEEE} {International} {Conference} on {Agents} ({ICA})},
	author = {Yourdshahi, Elnaz Shafipour and Pinder, Thomas and Dhawan, Gauri and Marcolino, Leandro Soriano and Angelov, Plamen},
	year = {2018},
	pages = {44--49},
}

@misc{rahman2023generating,
      title={Generating Teammates for Training Robust Ad Hoc Teamwork Agents via Best-Response Diversity}, 
      author={Arrasy Rahman and Elliot Fosong and Ignacio Carlucho and Stefano V. Albrecht},
      year={2023},
      eprint={2207.14138},
      archivePrefix={arXiv},
      primaryClass={cs.LG}, 
}

@inproceedings{rahman2024minimum,
author = {Rahman, Muhammad and Cui, Jiaxun and Stone, Peter},
title = {Minimum coverage sets for training robust Ad Hoc teamwork agents},
year = {2024},
publisher = {AAAI Press},
booktitle = {Proceedings of the Thirty-Eighth AAAI Conference on Artificial Intelligence and Thirty-Sixth Conference on Innovative Applications of Artificial Intelligence and Fourteenth Symposium on Educational Advances in Artificial Intelligence},
articleno = {1954},
numpages = {8},
series = {AAAI'24/IAAI'24/EAAI'24}
}

@inproceedings{
    mccallum2023is,
    title={Is feedback all you need? Leveraging natural language feedback in goal-conditioned {RL}},
    author={Sabrina McCallum and Max Taylor-Davies and Stefano Albrecht and Alessandro Suglia},
    booktitle={NeurIPS 2023 Workshop on Goal-Conditioned Reinforcement Learning},
    year={2023},
}

\clearpage

\appendix

\section{Architectural details}
All policy and value networks (including both low- and high-level for GRILL and GRILL-M) used hidden layer sizes of (64, 64). Additional dimensions were as follows:

\begin{table}[h]
    \centering
    \begin{tabular}{c|c|c}
         method & description & value \\
         \hline\hline
         LIAM & encoder hidden & (64, 64) (LSTM, dense) \\
         \hline
         LIAM & encoder output & 10 \\
         \hline
         LIAM & decoder hidden & (64, 64) \\
         \hline
         OMG & MLP hidden & 128 \\
         \hline
         OMG & MLP output & 32 \\
         \hline
         OMG & RNN hidden & (128, 128) \\
         \hline
         OMG & RNN output & 96 \\
         \hline
         OMG & VAE hidden & (128, 128) \\
         \hline
         OMG & VAE output & 16 \\
         \hline
         OMG & CVAE hidden & (128, 128) \\
         \hline
         OMG & CVAE output & 16 \\
         \hline
         GRILL-M & encoder hidden & (64, 64) (LSTM, dense) \\
         \hline
         GRILL-M & encoder output & 10 \\
         \hline
         GRILL-M & decoder hidden & (64, 64) \\
    \end{tabular}
    \label{tab:placeholder}
\end{table}

The `swish` activation function was used for all policy and value networks; all other networks used ReLU.

\section{Hyperparameters}

The following hyperparameters were used for the PPO backbone common to all methods (PPO, LIAM, OMG, GRILL, GRILL-M):

\begin{table}[h]
    \centering
    \begin{tabular}{c|c}
         hyperparameter & value  \\
         \hline\hline
         batch size & 1024 \\
         \hline
         minibatches & 8 \\
         \hline
         epochs & 2 \\
         \hline
         max grad norm & 0.5 \\
         \hline
         learning rate & 1e-4 \\
         \hline
         discount factor ($\gamma$) & 0.99 \\
         \hline
         GAE $\lambda$ & 0.95 \\
         \hline
         clipping parameter ($\epsilon$) & 0.2 \\
         \hline
         entropy coefficient & 0.01 \\
         \hline
         value loss coefficient & 0.5 \\
         \hline
         number of parallel envs & 16 \\
    \end{tabular}
    \label{tab:placeholder}
\end{table}

Additional method-specific hyperparameters were as follows:

\begin{table}[h]
    \centering
    \begin{tabular}{c|c|c}
         method & hyperparameter & value \\
         \hline\hline
         LIAM & encoder-decoder learning rate & 1e-3 \\
         \hline
         OMG & RNN, VAE, CVAE learning rate & 1e-3 \\
         \hline 
         GRILL-M & encoder-decoder learning rate & 1e-3 \\
         
    \end{tabular}
    \label{tab:placeholder}
\end{table}

\section{Training details}
Training runs were executed on a single NVIDIA RT2080Ti GPU. Agents were trained for 1e7 timesteps on cooperative reaching and 5e7 timesteps on level-based foraging. We note that to equalise the total number of `experienced' timesteps, we trained GRILL and GRILL-M online for 8e6 and 4.8e7 timesteps, since both methods used a dataset of 2e6 timesteps during the offline imitation phase. Training was repeated over 20 independent seeds (0-19) per method/environment pair (for a total of 200 training runs). Evaluation of each trained policy was performed over 1000 episodes (with unique seeds) for each of the three overlap scenarios. 

\end{document}